\documentclass[aps, twocolumn, prl]{revtex4-2}
\usepackage[latin1]{inputenc}
\usepackage{amsfonts, amssymb, braket, mathtools}
\usepackage[caption=false]{subfig}
\usepackage[export]{adjustbox}
\usepackage{subfloat}
\usepackage[colorlinks=true, linkcolor=red , citecolor=blue]{hyperref}
\usepackage[capitalise]{cleveref}
\usepackage{graphicx}
\usepackage[inline, shortlabels]{enumitem}
\usepackage{bbm}
\usepackage{bbold}
\usepackage{array}
\usepackage[normalem]{ulem}
\usepackage{color}
\usepackage{multirow}
\usepackage{xfp}
\usepackage{placeins}

\newcolumntype{C}{>{\centering\let\newline\\\arraybackslash\hspace{0pt}}m{9mm}}

\begin{document}

\title{Bell-state measurement exceeding 50\% success probability with linear optics}
\author{Matthias J. Bayerbach$^{1,2,\dagger}$, Simone E. D'Aurelio$^{1,2,\dagger}$, Peter van Loock$^3$, and Stefanie Barz$^{1,2}$}
\affiliation{
$^{1}$Institute for Functional Matter and Quantum Technologies, University of Stuttgart, 70569 Stuttgart, Germany\\
$^{2}$Center for Integrated Quantum Science and Technology (IQST), University of Stuttgart, 70569 Stuttgart, Germany\\
$^{3}$Johannes-Gutenberg University of Mainz, Institute of Physics, Staudingerweg 7, 55128 Mainz, Germany\\
$^\dagger$These authors contributed equally to this work.}

\begin{abstract}
Bell-state projections serve as a fundamental basis for most quantum communication and computing protocols today. However, with current Bell-state measurement schemes based on linear optics, only two of four Bell states can be identified, which means that the maximum success probability of this vital step cannot exceed 50$\%$. 
Here, we experimentally demonstrate a scheme that amends the original measurement with additional modes in the form of ancillary photons, which leads to a more complex measurement pattern, and ultimately a higher success probability of 62.5$\%$. Experimentally, we achieve a success probability of \((57.9 \pm 1.4) \%\), a significant improvement over the conventional scheme.
With the possibility of extending the protocol to a larger number of ancillary photons, our work paves the way towards more efficient realisations of quantum technologies based on Bell-state measurements.
\end{abstract}
 
\maketitle

\section{Introduction}
One of the most iconic applications of a Bell-state measurement (BSM) is quantum teleportation, where an arbitrary and unknown quantum state is teleported from one party to another by sharing entanglement between the parties and subsequently performing a BSM on the state to be teleported and one half of the entangled pair~\cite{Bennett1993, Gottesman1999, Braunstein1995, Kim2001,Barrett2004, Riebe2004, Lou2019, Takeda2013}. A BSM here means a projection onto a maximally entangled basis, the Bell basis. 

Quantum teleportation and in turn BSMs themselves now serve as important primitives for many other protocols underlying quantum technologies. In particular, in quantum communication, BSMs facilitate entanglement swapping and thereby the implementation of quantum repeaters. Furthermore, BSMs enable the realisation of measurement-device-independent quantum communication~\cite{Lo2012, Braunstein2012, Ma2018, Cui2019, Wei2020, Zhou2020, Nadlinger2021,Zhang2022}.
In quantum computing, BSMs have an important role in photonic quantum computing, in particular in measurement-based and fusion-based approaches~\cite{Raussendorf2001, Raussendorf2003,  Brown2005,  Segovia2015}, where, BSMs are an integral part of the generation of resource states for photonic quantum computing and in fusing small-scale units to large resource states for the realisation of quantum error correction ~\cite{Ewert2017, Varnava2007}.
Moreover, BSMs can be used to link distant stationary quantum computers or quantum registers via optical channels, thus creating a quantum internet~\cite{Lee2019, Hasegawa2019, Valivarthi2020, Azuma2015, Ewert2016, Pant2017}.

The standard approach to realising an optical BSM is letting two entangled photons impinge on two inputs of a balanced beam splitter and measuring the resulting output patterns and statistics ~\cite{Braunstein1995, Michler1996}. The simplicity of such a linear-optical approach comes at the cost of being able to identify only two out of four possible Bell states. This means that in $50\%$ of all cases, the obtained results are ambiguous ~\cite{Calsamiglia2001}. This limitation directly affects any optical quantum technology that relies on successful projection onto the Bell basis.

In atomic systems, complete BSMs  have been performed ~\cite{Riebe2004, Barrett2004, Gonzalez2019, Welte2021}. However, these require intricate experimental set-ups that are challenging to scale up. Complete BSMs on spin qubits are possible in solid-state systems~\cite{Reyes2022}, but thermal effects typically prevent operations at room temperature. Generally, non-optical approaches suffer from limited intrinsic clock rates of the order of MHz. Only the photonics platform offers, in principle, high processing clock rates at room temperature. Optical, continuous-variable Bell measurements and quantum teleportation~\cite{Takeda2013} are deterministic and well scalable, however, a practical mechanism for loss detection and general quantum error correction is lacking.

For the single-photon-based qubit approach, both scalability and fault tolerance in quantum communication and computing heavily depend on the BSM efficiencies. In this case, the $50\%$ limit has been overcome in proof-of-principle experiments by exploiting hyperentanglement ~\cite{Schuck2006, Barbieri2007, Li2017} and incorporating nonlinear elements ~\cite{Kim2001, Welte2021, Barrett2004, Kwiat1998}. A third approach that has been suggested in theory is based on adding additional ancillary photons to a linear-optics setup~\cite{Grice2011,Ewert2014} and using photon-number resolving detectors, offering clear advantages regarding lower experimental complexity and higher scalability. 

In this work, we demonstrate such a linear-optical BSM scheme enhanced by ancillary photons by adapting and implementing a scheme proposed in~\cite{Ewert2014}. In our experiments, we surpass the fundamental limit of $50\%$ demonstrating an experimental success probability of $57.9\%$. The maximum theoretical value for our approach is $5/8$~\cite{Ewert2014, Olivo2018}. 
We achieve this enhanced success probability by using additional single photons, expanding of the linear-optical circuits, and realizing pseudo-photon-number resolution using 48 single-photon detectors. Such an improved BSM efficiency can have a significant impact on practical quantum technologies and examples will be discussed.

In particular, these primitives are easily compatible with optical quantum computing and quantum networks and our results thus present a key step towards highly efficient Bell measurements for quantum technology.

\begin{figure*}[ht]%
    \centering
	\includegraphics[width = \linewidth, keepaspectratio]{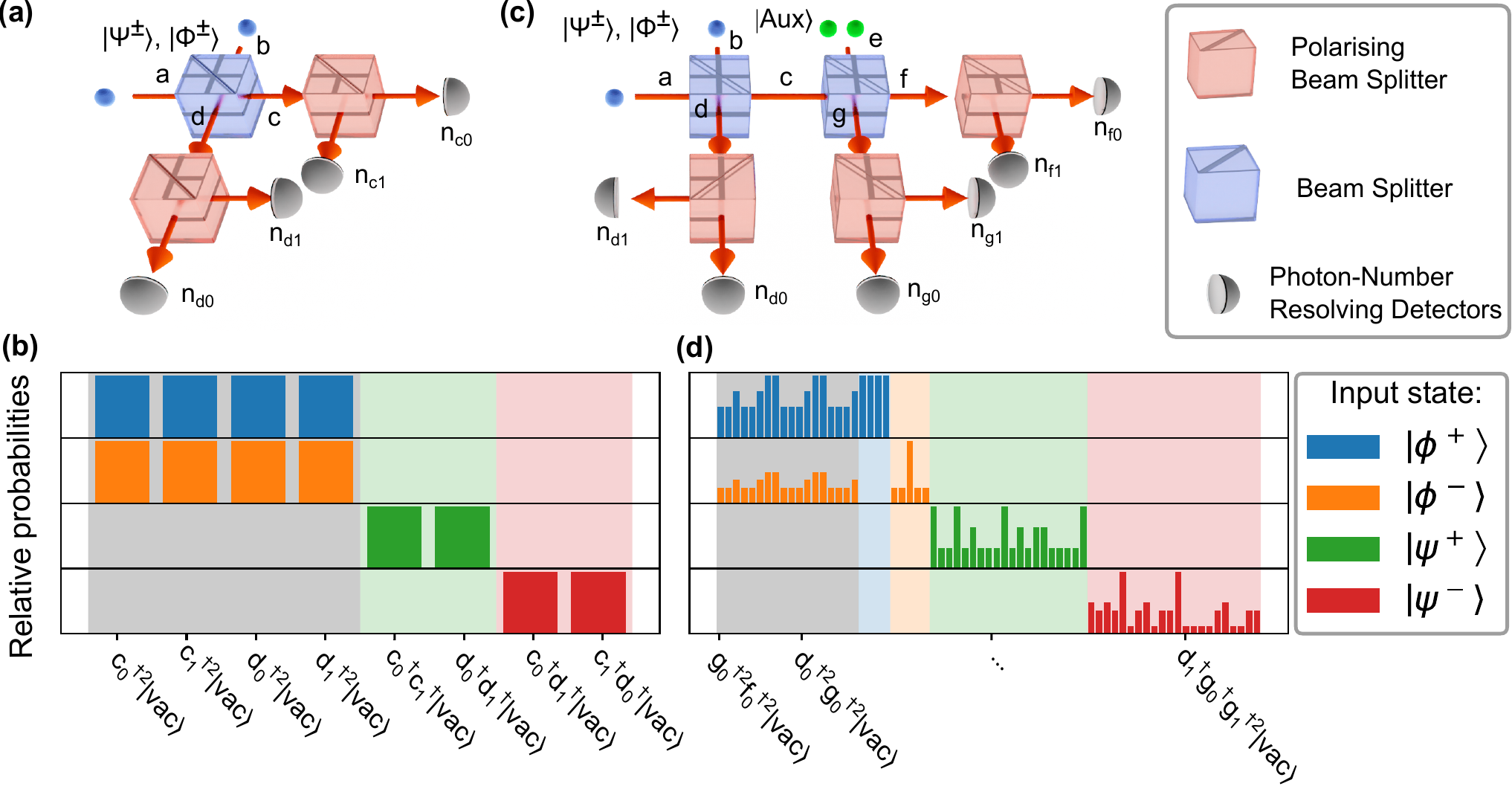}
	\caption{
    \textbf{(a)} Schematic of a standard BSM and \textbf{(b)} a BSM enhanced with an auxiliary state $\ket{Aux} = \frac{1}{2} (e^\dagger_0 e^\dagger_0+ e^\dagger_1 e^\dagger_1) \ket{vac}$, as well as their respective input-dependent detection-probability distributions \textbf{(c)} and \textbf{(d)}. Different input states are coloured and identifiable detection patterns are highlighted, while ambiguous ones are accentuated in gray. For the enhanced BSM, the setup is extended by a second beam splitter. Upon comparing the detection probability distributions, the advantage of the enhanced BSM becomes clear, as the two $\Phi^\pm$ states are now in $25\%$ of the cases identifiable.}
		\label{fig:Figure1}
\end{figure*}

\section{Theory} 

The Bell states are maximally entangled two-qubit states and are given by:
\begin{align}
\label{eq:Bellstates1}
\ket{\Psi^{\pm}} &= \frac{1}{\sqrt{2}} \left( \ket{\text{0}\text{1}}_{ab} \pm \ket{\text{1}\text{0}}_{ab} \right) 
			= \frac{1}{\sqrt{2}} \left( a^\dagger_0 b^\dagger_1 \pm a^\dagger_1 b^\dagger_0 \right) \ket{vac}, \\
\label{eq:Bellstates2}
\ket{\Phi^{\pm}} &= \frac{1}{\sqrt{2}} \left( \ket{\text{0}\text{0}}_{ab} \pm \ket{\text{1}\text{1}}_{ab} \right) 
			= \frac{1}{\sqrt{2}} \left( a^\dagger_0 b^\dagger_0 \pm a^\dagger_1 b^\dagger_1 \right) \ket{vac}.
\end{align}

Here, $\ket{0}$ and $\ket{1}$ denote the logical states of the two qubits $a$ and $b$, respectively. We can rewrite these terms in the form of the creation operators $a^\dagger_{0,1}$ and $b^\dagger_{0,1}$, which denote the creation of a photon in the spatial mode $a,\ b$ in the state 0 or 1 (for instance, encoded in two polarisation modes, $\ket{0}\equiv \ket{H}$, $\ket{1} \equiv \ket{V}$). The set of Bell states forms a basis of the two-qubit Hilbert space, the Bell basis.

In an ideal BSM, when inputting one of the four Bell states, we would identify that specific Bell state with $100\%$ probability. For optical qubits, BSMs can be realised with a setup consisting of a beam splitter \cite{Braunstein1995}.
The Bell state impinging on a balanced beam splitter (see Fig. \ref{fig:Figure1}) transforms into the following output states:
\begin{align}
\ket{\Psi^{+}}      \rightarrow& \frac{1}{\sqrt{2}} \left( c^\dagger_0 c^\dagger_1 +  d^\dagger_0 d^\dagger_1\right) \ket{vac}, \\
\ket{\Psi^{-}}      \rightarrow& \frac{1}{\sqrt{2}} \left(c^\dagger_0 d^\dagger_1- c^\dagger_1 d^\dagger_0 \right)\ket{vac}, \\
\ket{\Phi^{\pm}}    \rightarrow& \frac{1}{2\sqrt{2}} ( c^\dagger_0 c^\dagger_0 \pm c^\dagger_1 c^\dagger_1+ d^\dagger_0 d^\dagger_0 \pm d^\dagger_1 d^\dagger_1 ) \ket{vac} ,
\end{align} 
where $c$ and $d$ are the output modes of the beam splitter (see Fig. \ref{fig:Figure1}\textbf{(a)}). 
Measuring the output according to the individual spatial modes and qubit states $\{0,1\}$ results in a distinctive photon pattern for the states $\ket{\Psi^\pm}$. However, we see that the states $\ket{\Phi^\pm}$ cannot be uniquely identified (Fig. \ref{fig:Figure1}\textbf{(b)}). Note that linear-optical setups can be designed to distinguish between different sets of states, but the total success probability of the unambiguous output patterns combined never exceeds one half, imposing the $50\%$ limit ~\cite{Calsamiglia2001}. 

We will now show how this probability can be improved using additional ancillary photons in the state (see also~\cite{Ewert2014}): 
\begin{equation}
\ket{Aux} = \frac{1}{2} (e^\dagger_0 e^\dagger_0+ e^\dagger_1 e^\dagger_1) \ket{vac}. 
\end{equation}
We use another beam splitter (Fig. ~\ref{fig:Figure1}\textbf{(c)}) and send the ancillary state into one of the inputs of this second beam splitter (mode $e$). We send one output mode $c$ from the first beam splitter to the other input of this second beam splitter. After this second beam splitter, we obtain an output state that consists of four photons distributed over the modes $d$, $f$, and $g$ (see Fig. \ref{fig:Figure1}\textbf{(c)}). We observe a distinct photon-number distribution for each Bell state as shown in Fig. \ref{fig:Figure1}\textbf{(d)}.

As before in the standard scheme, we can still distinguish the states $\ket{\Psi^\pm}$ with $100~\%$ probability. However, we get now in addition a subset of unique signatures for the states $\ket{\Phi^+}$ and $\ket{\Phi^-}$ in $25\%$ of the cases (see Fig. \ref{fig:Figure1} \textbf{(d)}).
This means that the overall probability to correctly identify a Bell state for this scheme is $p_{c} = 62.5~\%$~\cite{Ewert2014}. 
At the cost of more ancillary photons, this scheme can be expanded to reach success probabilities close to unity~\cite{Ewert2014,Grice2011}. This protocol can be summarized as a black box that takes one of the Bell states as input, and outputs a label that can be either one of the four Bell states or an inconclusive result.
\begin{figure*}%
    \centering
	\includegraphics[width=.99\textwidth]{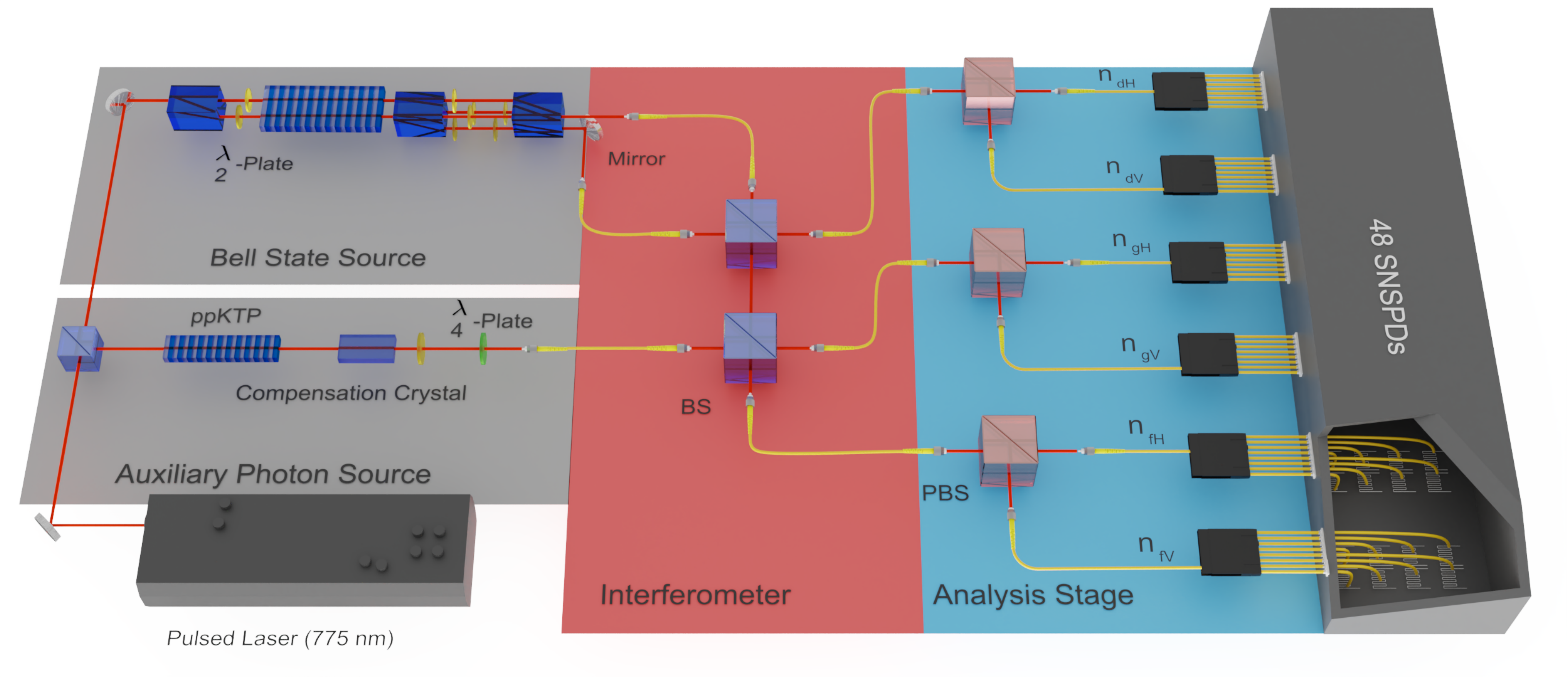}
	\caption{
	Schematic representation of an SPDC-based enhanced BSM, using PPKTP crystals. The setup consists of a Bell-state and auxiliary state source (left, grey area), an enhanced BSM interferometer (middle, red area) and an analysis stage (right, blue area). The auxiliary source uses a PPKTP for photon pair generation, a compensation crystal to counteract the resulting temporal walk-off, and waveplates to set the correct state. For the Bell-state source, a PPKTP crystal is placed inside a Mach-Zehnder-like interferometer. The generated Bell-state photons enter the upper beam splitter, one of which's output is sent to the lower beam splitter, together with the auxiliary state. The three output modes are then split by PBSs resulting in six spatial modes, which are each again split into eight for detection purposes.
    This leaves 48 detection modes each of which are connected to a superconducting nanowire detector. Registered detection events are analyzed by a timetagger system and the number of photons in each spatial mode is computed.}
	\label{figure:Setup}
\end{figure*}

We now define a number of parameters that will allow us to quantify the success of the BSM. 
First, we assume we send in a particular Bell state. In general, a measurement outcome can either be \textit{unambiguous} or \textit{ambiguous}, meaning the pattern does or does not allow for the identification of a specific Bell state.
Within the unambiguous results, we can then have two cases: either the BSM outputs the correct label or it does not.

Thus, the first quantity is the \textit{probability of an unambiguous and correct result} $p_c$: the probability to obtain an unambiguous and correct result given a certain Bell state as input and a measurement outcome $m$, averaged over all possible Bell states,
\begin{equation} 
    p_c = \frac{1}{4}\sum_{\ket{\psi} \in \left\{\ket{\Psi^\pm},\ket{\Phi^\pm}\right\}}  P\left(m=\psi\bigg\vert\ket{\psi}\right).
\end{equation}
This probability can reach up to $62.5\%$ in the case of our enhanced scheme.

Due to experimental imperfections, the scheme might output an unambiguous result, which is not consistent with the input state. The probability for \textit{unambiguous and false measurement} $p_f$ is given by:
\begin{equation}
    p_f = \frac{1}{4}\sum_{\ket{\psi} \in \left\{\ket{\Psi^\pm},\ket{\Phi^\pm}\right\}} P\left(m\neq\psi\bigg\vert\ket{\psi}\right),
\end{equation}
and we have $p_c+p_f=1-p_{amb}$, where $p_{amb}$ is the probability for an ambiguous measurement.

Knowing $p_c$ and $p_f$, allows us to derive the measurement discrimination fidelity, defined as~\cite{Wein2016}:
\begin{equation}
    \mbox{MDF} = \frac{p_c}{p_c+p_f}.
\label{eq:MDF}
\end{equation}
The $\mbox{MDF}$ denotes the percentage of all correct labels in the subset of all unambiguous results and lies between zero (all labels are incorrect) and one (all labels are correct).

Finally, to analyse the complete output statistics, we use the total variation distance $D$ \cite{Wang2019}:
\begin{equation}
    D=\sum\frac{\left|f_i-q_i\right|}{2},
    \label{eq:distance}
\end{equation}
where \(q_i\) is the expected theoretical probability, while \(f_i\) is the measured relative probability for the \textit{i}-th four-photon output pattern.

\begin{figure*}
    \centering
    \includegraphics[width = 0.99\linewidth]{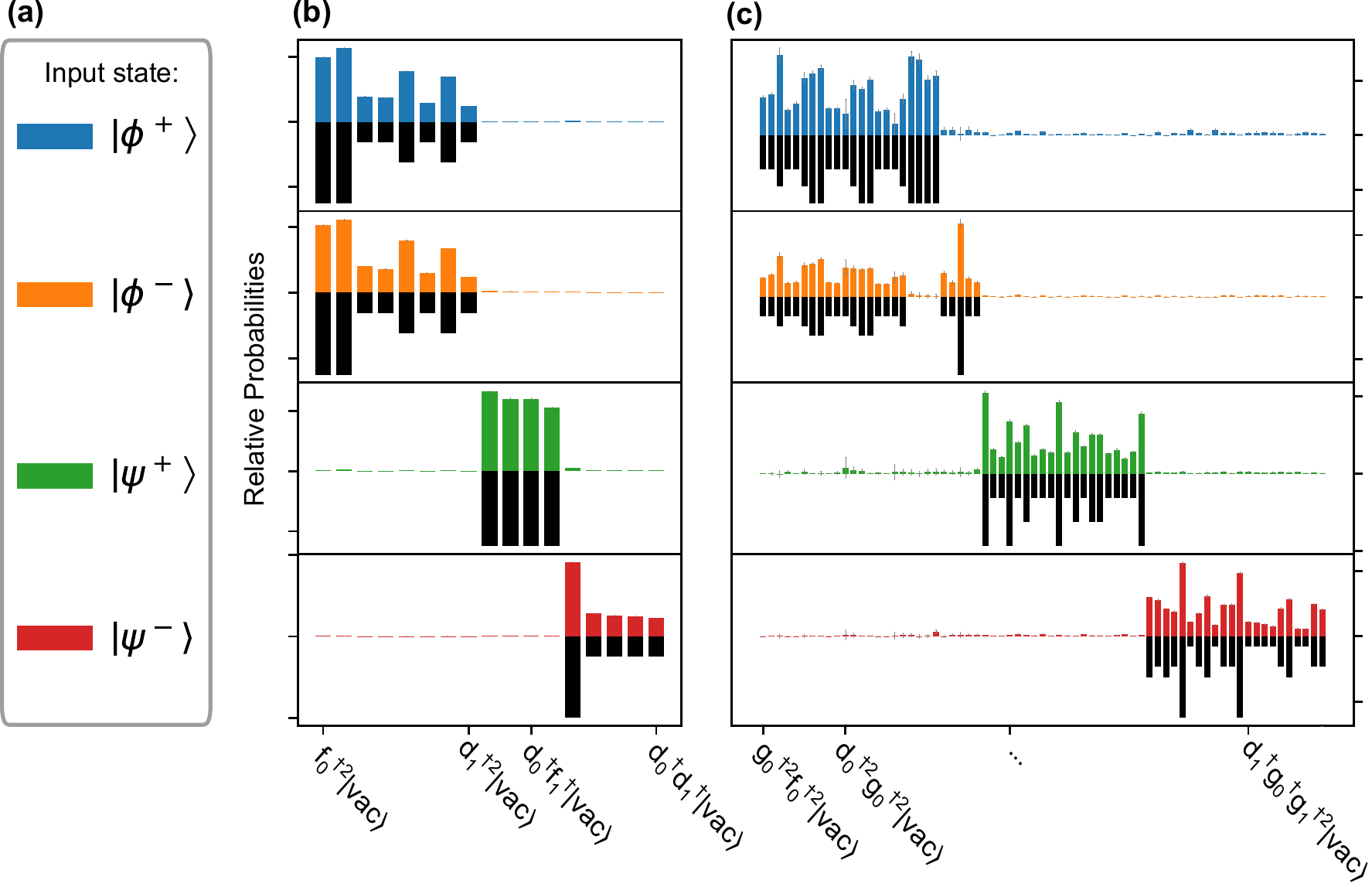}
    \caption{
    The figure shows the relative probabilities of the detection patterns, \textbf{(a)} for each input state, \textbf{(b)} for standard and \textbf{(c)} enhanced BSM. 
    For comparison, theoretical values are represented by black bars and are mirrored with respect to zero. Error bars are shown in grey. The labels represent the photon number for each mode (Eq.\eqref{eq:OutputBasis}). 
    While the standard BSM is showing some variations due to experimental imperfections, the measured data are in agreement with the expectations, as captured by the average distance of $D = 0.072$. A standard probability of a correct result of $p_c=0.481$ was achieved.
    The data of the enhanced BSM also matches the theoretical data well and the enhanced probability has increased to $p_c = 0.579$.}
    \label{fig:Figure4}
\end{figure*}

\section{Experiment}
Whereas the original proposal was based on information being encoded in the path of single photons~\cite{Ewert2014}, in our implementation, we use polarisation as a degree of freedom: $a^\dagger _0 :=a^\dagger _H$ and $a^\dagger _1:=a^\dagger _V$. Note that the scheme can be adapted to arbitrary encodings of photonic qubits. 
We generate both photonic Bell states and the ancillary states by using  parametric down-conversion  in periodically poled potasium titanyl phosphate (ppKTP) crystals (see Fig. ~\ref{figure:Setup}).
For the generation of the different Bell states (see Eqn. (\ref{eq:Bellstates1},~\ref{eq:Bellstates2})), we use a collinear arrangement that generates polarisation-entangled states using a Mach-Zehnder-type interferometer~\cite{Evans2010}. We characterise the quality of the Bell states by performing correlation measurements, both in the basis $\{H,V\}$ and the basis $\{+,-\}$ with $\ket{+}= 1/\sqrt{2}(a^\dagger_H + a^\dagger_V)\ket{vac}$ and  $\ket{-}= 1/\sqrt{2}(a^\dagger_H - a^\dagger_V)\ket{vac}$, obtaining visibilities of $V_{H/V}=0.975$ and $V_{+/-}=0.954$.
The auxiliary state is obtained with a second collinear setup. The photons are generated from the crystal in the product state $e^\dagger_H e^\dagger_V \ket{vac}$ and, with the use of a half-waveplate and a quarter-waveplate at respective angles $\theta_{HWP}=22.5^\circ$ and $\theta_{QWP}=45^\circ$, the final state becomes $\ket{Aux}=\frac{1}{2} (e^\dagger_He^\dagger_H+e^\dagger_Ve^\dagger_V)\ket{vac}$. We measure the visibility of the auxiliary state in the basis $\{H,V\}$, achieving a value of $V_{H/V}=0.9899$

The apparatus for the BSM is composed of two balanced beam splitters with the outputs being coupled into single-mode fibres. These route the photons to an analysis stage, which allows performing polarisation measurements using polarising beam splitters (PBSs). The six output modes of this stage will be labeled as $d_{H,V},\ f_{H,V}, \ g_{H,V}$ (Fig. \ref{figure:Setup}). The final state is defined in a Hilbert space described by the following set of basis vectors:
\begin{equation}
\label{eq:OutputBasis}
\left\{ \ket{n_{dH} n_{dV} n_{fH} n_{fV} n_{gH} n_{gV}} \bigg\vert \sum_i n_i = 4\right\},
\end{equation}
where $n$ indicates the photon number of each particular mode.
Each of these modes are further split up into eight spatial modes using a fibre-based $1\times8$ splitter, allowing pseudo-photon-number resolution. The photons are detected using superconducting nanowire single-photon detectors (SNSPD) with an detection efficiency of on average $88.6\%$.

\begin{figure*}
    \centering
      \includegraphics[width=.99\textwidth]{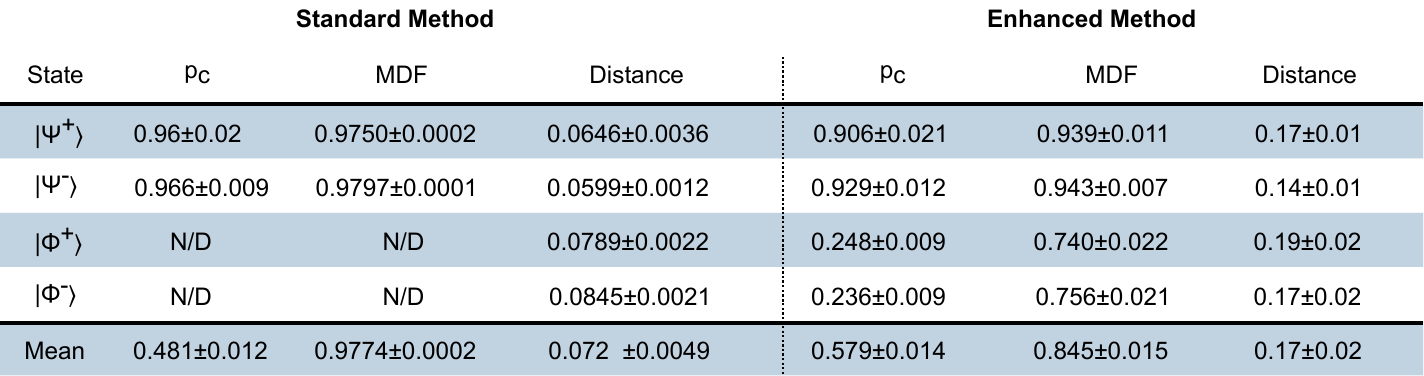}
	\caption{Measured quantities for both the normal and enhanced method for each of the four Bell states. For the $\ket{\Phi^\pm}$ states, the measurement fidelity is not defined in the normal scheme, since the theory does not allow any unambiguous pattern.}
	\label{tab:fidelity}
\end{figure*}
To compare both BSM methods, two sets of measurements are performed for each Bell state $\left\{\ket{\Psi^\pm},\ket{\Phi^\pm}\right\}$. The first one is the standard approach, in which the photons from the Bell state go through the apparatus without the presence of the auxiliary state, recording the output statistics. 
In this case, the presence of the second beam splitter increases the number of possible output patterns, but does not affect the probability to correctly identify a Bell state.
Then, the enhanced BSM is tested by switching the ancillary source on and, again, recording the output statistics.
We take into account higher-order emission and induced coherence~\cite{Ou1990} by measuring those contributions for each source separately and subtracting those counts from the signal in the postprocessing.
Finally, the raw count rates are corrected by a factor introduced by the probabilistic photon number detection (see Appendix) in order to calculate the previously introduced quantities starting from this data.

\section{Results}

We first measure the average probability of the standard BSM. Our results are shown in Fig. ~\ref{fig:Figure4} \textbf{(a)}. 
The standard scheme allows us to identify both $\ket{\Psi^\pm}$ states with over $96\%$ probability, while the $\ket{\Phi^\pm}$ states are completely ambiguous. The measured distributions result in a probability of $(48.1\pm 1.2)\%$ to correctly identify the incoming state. As can be seen from Table \ref{tab:fidelity}, we achieve high values for parameter MDF ($0.977 \pm 0.0002$) and low values for the distance D ($0.072 \pm 0.0049$), which indicate the general high quality of optical BSMs. However, our results are close to the theoretical upper bound of $50\%$, limiting the possibility to increase $p_c$.

As a next step, we demonstrate the enhanced measurement protocol (see Fig. ~\ref{fig:Figure4} \textbf{(b)}). Whilst in the standard BSM the states $\ket{\Phi^\pm}$ do not show a unique pattern, the enhanced BSM features a subset of distinguishable outcomes allowing the identification of these states. From the data, we can estimate an average probability of an unambiguous and correct result, with $(57.9 \pm 1.4)\%$. In contrast to the previous measurement, we can now identify more than $23 \%$ of $\ket{\Phi^\pm}$ states, while keeping $p_c$ for the $\ket{\Psi^\pm}$ states above $0.9$. This increases the total probability of a correct identification above the theoretical limit of $50\%$.
A full list  of characterisation parameters for each state is shown in Table~\ref{tab:fidelity}.

\begin{figure}
    \centering
    \includegraphics[width = 0.99\linewidth]{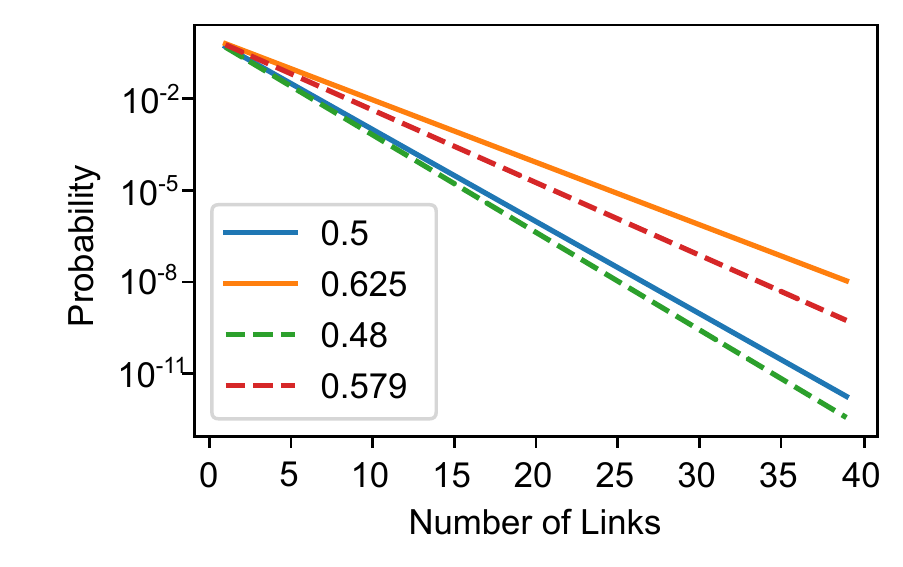}
    \caption{Probability of transmission for a memoryless quantum relay using entanglement swapping. Different lines represent different BSM success probabilities. Solid lines indicate theoretical values while dashed lines indicate the experimental values obtained in this work for the standard and the enhanced BSM schemes, respectively.}
    \label{fig:Figure5}
\end{figure}

In Fig.~\ref{fig:Figure5}, we illustrate the implications of improving the BSM success probability from $50\%$ to $57.9\%$ for a quantum relay that connects the segments of a quantum channel via entanglement swapping at intermediate stations. Such a BSM-based relay achieves, for instance, long-distance privacy. Due to the exponential scaling of the overall success probability with the number of segments $n$, $p_c^{n-1}$, already a slightly enhanced swapping efficiency has already a significant impact. While this does not improve the photon-loss scaling in a real fiber channel compared with a point-to-point fiber link, detecting the presence of the photons at the intermediate stations can be beneficial with dark counts. Note that when the stations are equipped with quantum memories in a quantum repeater chain, the loss scaling is typically improved to $p_c^{log_2(n)}$ in certain regimes and settings.

A much more significant, practical gain can be obtained for an all-photonic memoryless quantum repeater combining many physical qubits into a few logical qubits of a suitable quantum error correction code. For codes being based on single-photon sources and probabilistic linear-optics fusion gates using BSMs, the code's state generation can be made much more efficient. In particular, the enhancement we achieved in our experimental scheme of $58\%$ compared to the standard $50\%$ leads to a significant reduction of the overhead.

Let us consider the scheme proposed in~\cite{Ewert2017} with a memoryless repeater every 2km over a total distance of 1000km. The scheme is based on logical qubits being encoded in $n$ blocks each containing $m$ photons.
When using a small code, i.e. 4 blocks with 2 photons per block, the total average number of photons per node is reduced by a factor of 2 for a BSM success probability of $58\%$ compared to $50\%$. For a larger code, i.e. 67 blocks with 11 photons per block, we obtain already a reduction of a factor of 5 per node. In total, this allows achieving long-distance distribution success probabilities greater than $50\%$ with a significant reduction of the overhead. Further improvement to $62.5\%$ or even $75\%$ in the BSM success probabilities would reduce the overhead by one or even several orders of magnitude, respectively.

The significant impact of the enhanced BSM on the state generation, and thus also on the total rates,
of an all-photonic quantum repeater based on photonic cluster states~\cite{Azuma2015}, leading to an overhead reduction of several orders of magnitude, was also found in~\cite{Pant2017}.

\section{Conclusions}

We report the first implementation of a BSM on photonic qubits with a success probability surpassing the $50\%$ limit using only linear optics and two ancillary photons.
In our experiments we reach a success probability $p_c=57.9 \%$, while a standard BSM, measured under the same experimental conditions, only achieved $p_c=48\%$.
For this class of BSMs, the two ancillary photons represent the minimal extra resource required to beat that notorious $50\%$ bound, i.e. there is evidence that adding just a single ancillary photon is insufficient~\cite{Olivo2018}.

We note that our implementation does not require a deterministic auxiliary-state source; even when the source fails to produce a state, BSMs with a $50\%$ success rate are still possible with this setup. This is due to the underlying nature of the scheme to only increase and never reduce any given success probability.

Looking ahead, even higher success rates could be achieved by extending the measurement scheme: the addition of a second ancillary photon pair or  $\ket{\Phi^-}$ as an auxiliary state would, in principle, boost the maximum success rate to $75\%$. By further scaling up the setup and adding more ancillary states, the success rate of the measurement can be enhanced arbitrarily close to $100 \%$ as shown in ~\cite{Grice2011,Ewert2014}.

Our results demonstrate how ancillary photon states and linear-optical setups can improve the success rates of BSMs, offering a viable option to boost the general efficiency of any quantum protocol using BSMs.

One potential future application could be the creation of large cluster states for measurement-based quantum computing, as it would be possible, applying concepts of percolation theory, to create cluster states of the required size, once the $62.5\%$ efficiency threshold is surpassed with the proper resource states~\cite{Segovia2015,Zaidi2015}. 

Combined with the significance of BSM efficiency for the field of quantum communication in particular, this work could serve as a stepping stone towards larger quantum networks and more-efficient communication links in the future.

\section{Acknowledgments}
We thank Helen Chrzanowski, Shreya Kumar, Nico Hauser, Daniel Bhatti, and David Canning for insightful discussions and helpful comments. 
We would like to acknowledge support from the Carl Zeiss Foundation, the Centre for Integrated Quantum Science and Technology (IQ$^\text{ST}$), the German Research Foundation (DFG), the Federal Ministry of Education and Research (BMBF, project SiSiQ and PhotonQ), and the Federal Ministry for Economic Affairs and Energy (BMWi, project PlanQK). P.v.L. also acknowledges support from the BMBF via QR.X and from the BMBF/EU via QuantERA/ShoQC.

\newpage
\onecolumngrid 
\appendix

\section{Appendix}

\subsection{A) Complete Output States}
The output states in the modes $f$ and $g$ are listed below. We only consider the states $\ket{\Psi^+}$ and $\ket{\Phi^\pm}$ as input, since the total photon number in both modes is four. The state $\ket{\Psi^-}$ has only three photons in these two modes and can therefore always be identified by its photon number.
\begin{align*}
\label{eq:completestate}
\ket{\Psi^{+}} \ket{Aux}  \rightarrow  \frac{1}{8}&\left( - a^{\dagger 3}_{f0}a^{\dagger}_{f1} + i a^{\dagger 2}_{f0}a^{\dagger}_{f1}a^{\dagger}_{g0}  - a^{\dagger}_{f0}a^{\dagger}_{f1}a^{\dagger 2}_{g0} +i a^{\dagger}_{f1}a^{\dagger 3}_{g0} -i a^{\dagger 3}_{f0}a^{\dagger}_{g1} \right. \\
& \left. - a^{\dagger 2}_{f0}a^{\dagger}_{g0}a^{\dagger}_{g1} -i a^{\dagger}_{f0} a^{\dagger 2}_{g0} a^{\dagger}_{g1} 
- a^{\dagger 3}_{g0} a^{\dagger}_{g1} - a^{\dagger}_{f0} a^{\dagger 3}_{f1}  \right. \\
& \left. +i a^{\dagger}_{g0} a^{\dagger 2}_{f1} a^{\dagger}_{g1} - a^{\dagger}_{f0} a^{\dagger}_{f1} a^{\dagger 2}_{g1}  + i a^{\dagger}_{f0} a^{\dagger 3}_{g1} - i a^{\dagger 3}_{f1} a^{\dagger}_{g0} -  a^{\dagger 2}_{f1} a^{\dagger}_{g0} a^{\dagger}_{g1} \right. \\
& \left.- i a^{\dagger}_{f1} a^{\dagger}_{g0} a^{\dagger 2}_{g1} -  a^{\dagger}_{g0} a^{\dagger 3}_{g1} \right) \ket{vac} \\ 
\ket{\Phi^{+}} \ket{Aux}    \rightarrow& \frac{1}{8} \left[ \textcolor{blue}{-\frac{1}{2} \bigg( a^{\dagger 4}_{f0} + a^{\dagger 4}_{f1} + a^{\dagger 4}_{g0}+ a^{\dagger 4}_{g1} \bigg) - a^{\dagger 2}_{f0} a^{\dagger 2}_{g0}  - a^{\dagger 2}_{f1} a^{\dagger 2}_{g1}} \right. \\
& \left. - a^{\dagger 2}_{f0} a^{\dagger 2}_{f1} + a^{\dagger 2}_{f0} a^{\dagger 2}_{f1} + a^{\dagger 2}_{f1} a^{\dagger 2}_{g0} - a^{\dagger 2}_{g0} a^{\dagger 2}_{g1} -4 a^{\dagger}_{f0} a^{\dagger}_{f1} a^{\dagger}_{g0} a^{\dagger}_{g1} \right] \ket{vac} \\
\ket{\Phi^{-}} \ket{Aux}   \rightarrow& \frac{1}{8} \left[ \textcolor{blue}{-\frac{1}{2} \bigg( a^{\dagger 4}_{f0} + a^{\dagger 4}_{f1} + a^{\dagger 4}_{g0}+ a^{\dagger 4}_{g1} \bigg)  - a^{\dagger 2}_{f0} a^{\dagger 2}_{g0} - a^{\dagger 2}_{f1} a^{\dagger 2}_{g1} } \right] \ket{vac} \\
& +\frac{i}{4}\bigg( a^{\dagger 2}_{f0} a^{\dagger}_{f1} a^{\dagger}_{g1} - a^{\dagger}_{f0} a^{\dagger 2}_{f1} a^{\dagger}_{g0} + a^{\dagger}_{f0} a^{\dagger}_{g0} a^{\dagger 2}_{g1} - a^{\dagger}_{f1} a^{\dagger 2}_{g0} a^{\dagger}_{g1} \bigg)   \ket{vac}
\end{align*}

The shared terms of the $\ket{\Phi^\pm}$ states are written in blue. The subscript indicates the output mode of the beam splitter.

\subsection{B) Pseudo-Photon-Number Resolving Detectors}
The detectors used to measure the photon states are superconducting nanowire single-photon detectors. While these detectors have very high detection efficiencies of more than $90\%$ for photons at 1550 nm and a very low dark count rate of less than 100 counts per second, the detectors are binary and cannot distinguish between single- and multi-photon events. To reveal the photon numbers nevertheless, each mode is equally split into 8 modes. Each of these modes connects to a superconducting nanowire detector. The number of coincidences between the detectors reveals the photon number of the original mode. This method is biased though, as multiple photons can be routed to the same output by the splitter. These events result in an incorrect photon number and are discarded in the post-selection. The probability for a correct measurement therefore depends on the number of detectors $k$ and the number of photons $n$ and can be written as:
\begin{equation*}
P(n,k) = \frac{k!}{(k-n)!}\frac{1}{k^{n}}
\end{equation*}
Since this value depends on the photon number, the measured statistics is shifted by this distribution. To correct for this effect, each measurement outcome of the form $\ket{n_{dH}n_{dV}n_{fH}n_{fV} n_{gH} n_{gV}}$ has to be multiplied by a factor $\frac{1}{P_{PPNR}}$:
\begin{equation*}
    P_{PPNR} = \prod_{i=1}^{6} P(n_i, k)
\end{equation*}
Here $k=8$, as each mode is split onto eight detectors.

\end{document}